\documentclass[aps,groupedaddress,nofootinbib,showkeys,showpacs,amsfonts,eqsecnum]{revtex4}

\usepackage{epsfig}
\usepackage{graphicx}
\usepackage{amsmath}

\newcommand{\A}{{{\mathcal{A}}}}

\newcommand{\tr}{{\rm tr}}

\newcommand{\bfk}{{{\boldsymbol{k}}}}

\newcommand{\bfx}{{{\boldsymbol{x}}}}

\newcommand{\thru}[1]{\mathrel{\mathop{#1\!\!\!\!/}}}

\begin{document}

\title{Dimension two condensates and the Polyakov loop above the
deconfinement phase transition}

\author{E. Meg\'{\i}as}
\email{emegias@ugr.es}

\author{E. \surname{Ruiz Arriola}}
\email{earriola@ugr.es}

\author{L. L. Salcedo}
\email{salcedo@ugr.es}

\affiliation{
Departamento de F\'{\i}sica Moderna,
Universidad de Granada,
E-18071 Granada, Spain}

\date{\today} 

\begin{abstract}
We show that recent available lattice data for the renormalized
Polyakov loop above the deconfinement phase transition exhibit
unequivocal inverse power temperature corrections driven by a
dimension 2 gluon condensate. This simple ansatz provides a good
overall description of the data throughout the deconfinement phase
until near the critical temperature with just two parameters. One of
the parameters is consistent with perturbation theory while a second,
non perturbative, parameter provides a numerical value of the
condensate which is close to existing zero and finite temperature
determinations.
\end{abstract}

\pacs{12.38.Mh,12.38.Gc,12.38.Aw}

\keywords{Lattice, Polyakov loop, Finite Temperature, Deconfinement,
Dimension-2 condensate}

\maketitle

\section{INTRODUCTION}
\label{sec:intro}

The Polyakov loop plays a relevant theoretical role in QCD at finite
temperature. It represents the propagator of a static test quark and
therefore it is crucial in the understanding of the
confinement-deconfinement crossover. In
\cite{Kuti:1980gh,McLerran:1981pb} it was related to a heavy quark
free energy so its vanishing in quenched QCD signals the confinement
phase. As noted by 't Hooft \cite{'tHooft:1977hy}, gluodynamics at
finite temperature formulated using the imaginary time formalism, has
an extra discrete global symmetry, in addition to usual gauge
invariance. This symmetry is spontaneously broken above the
deconfinement phase transition
 \cite{Polyakov:1978vu,Susskind:1979up}. The Polyakov loop, $L(T)$, is
a natural order parameter for such phase transition; under periodic
gauge transformations $L$ is an invariant object but under a 't Hooft
transformation it picks up a factor which is an element of the center
of the gauge group. Effective field theories for the Polyakov loop
have been proposed in \cite{Pisarski:2000eq}. (For a comprehensive
review see e.g. Ref. \cite{Pisarski:2002ji}).

The smooth Wilson loops, and in particular the Polyakov loop, are
composite operators. Their perturbative renormalizability was
discussed in
\cite{Polyakov:1980ca,Arefeva:1980zd,Dotsenko:1979wb,Gervais:1979fv},
finding the remarkable result that they are multiplicatively
renormalizable, without mixing with other operators. Soon afterwards,
the perturbative evaluation of the Polyakov loop was addressed by Gava
and Jengo \cite{Gava:1981qd} within dimensional regularization, to
next-to-leading order (NLO). After including finite temperature vacuum
polarization effects through Debye mass insertion, the leading order
term turns out to be ${\cal O}(g^3)$ instead of the naively expected
${\cal O}(g^2)$. Their result implies that at high enough temperatures
the renormalized Polyakov loop should approach unity from above, a
consequence of the non trivial factor introduced by the
renormalization. (The expectation value of the bare Polyakov loop
vanishes in the continuum limit in any phase.) Not much progress has
been achieved after this early result. At present there are no
perturbative calculations of the expectation value of the Polyakov
loop beyond NLO. As noted in \cite{Gava:1981qd} a direct calculation
would have to confront the proliferation of Feynman diagrams due to
infrared divergences \cite{Linde:1980ts}. A different approach,
related to the dimensional reduction technique, is discussed below.

On the non perturbative side, the bare Polyakov loop has often been
studied numerically within lattice gauge theory calculations, however,
a reliable definition and calculation of the renormalized Polyakov
loop has been achieved only recently. The method introduced in
Ref. \cite{Kaczmarek:2002mc} for quenched QCD obtains the Polyakov
loop as a byproduct of the heavy quark-antiquark potential at finite
temperature, obtained from the correlation between two Polyakov loops
at different separations. Comparison with the zero temperature
potential for small separations allows a quite precise determination
of the quark selfenergy to be removed and so of the Polyakov loop. The
renormalized Polyakov loop is larger than unity for temperatures at
and above $3\, T_c $, in agreement with the perturbative
expectation. The same technique has been applied to two flavor QCD in
\cite{Kaczmarek:2005ui}. A direct lattice calculation of the Polyakov
loop has also been reported in Ref. \cite{Dumitru:2003hp} using a
different approach. In this case a single Polyakov loop is
used. Comparison of data taken at different temperatures allows to
determine the renormalization factor to be applied to the bare
result. The results of these two approaches agree approximately near
the phase transition, but for temperatures above $1.3\,T_c$ the
behaviors turn significantly different. The differences could be due
to the effects of finite lattice spacing or to ambiguities in the
renormalization prescription.

High temperatures probe kinematical regions which up to the manifest
breaking of the Lorentz invariance correspond to large Euclidean
momenta in the zero temperature quantum field theory. In dimensional
regularization in the $\overline{\rm MS}$ scheme one finds that to a
given temperature $T$ there corresponds an Euclidean scale $\mu \sim 4
\pi T $ \cite{Huang:1994cu}, so that $T_c= 270\,$MeV means $\mu=
3\,$GeV. In this regime one expects Operator Product Expansion (OPE)
ideas to apply and more specifically, at not too high temperatures,
condensates and power corrections should play a role. Actually,
following some older proposals \cite{Lavelle:1988eg}, phenomenological
requirements \cite{Chetyrkin:1998yr}, theoretical
studies \cite{Kondo:2001nq} and lattice analyses
 \cite{Boucaud:2001st,RuizArriola:2004en,Boucaud:2005rm} there has
recently been mounting evidence that the lowest condensate order BRST
invariant condensate is of dimension 2. Such a condensate is generally
non-local but in the Landau gauge becomes the local operator $\langle
A_{\mu,a}^2\rangle $, with $A_{\mu,a} $ the gluon field. Also the
$\langle A_0^2\rangle$ condensate appears as a parameter in the
calculation of the pressure at finite temperature
\cite{Kajantie:2000iz}.

In this work we investigate the role of condensates on the expectation
value of the Polyakov loop. The Polyakov loop is closely related to
the thermal expectation value of $\tr(A_0^2)$ (the NLO perturbative
result can be obtained in this way) and so condensate contributions to
this quantity would have immediate impact on the Polyakov loop. Our
motivation is best exposed by drawing an analogy with the zero
temperature quark-antiquark potential in quenched QCD. The potential
is, of course, closely related to the correlation function of two
thermal Wilson lines. The perturbative regime of the potential $V(r)$
corresponds to small separations, where the potential is approximately
Coulombian. At separations of the order of $1/\Lambda_{\text{QCD}}$
(there is no other scale in gluodynamics) a linearly confining term
develops and starts becoming dominant. Both pieces of the potential
evolve under the renormalization group at a logarithmically slow
rate. Therefore, modulo radiative corrections, the dimensionless
quantity $rV(r)$ is composed of a flat perturbative piece plus a
power-like term of the type $\Lambda_{\text{QCD}}^2r^2$ which is non
perturbative. In analogy, at high temperatures, we can consider the
behavior of the dimensionless quantity $\langle
\tr(A_0^2)\rangle/T^2$, also directly related to the correlation
function of two thermal Wilson lines. The analogous of the scale $r$
in the previous case is the scale $1/T$ here, and certainly for large
$T$ the quantity $\langle \tr(A_0^2)\rangle/T^2$ is perturbative and
flat modulo a logarithmic dependence. At lower temperatures we
contemplate the possibility of non perturbative power-like terms of
the type $\Lambda_{\text{QCD}}^2/T^2$ to develop. As we show in this
work, such term enters naturally through OPE corrections to the gluon
propagator driven by condensates. An analysis of available lattice
data turns out to display precisely the power-like pattern expected
from the previous considerations. The pattern is followed in the
deconfinement phase from the highest temperatures available down to
near to the transition where deviations start to show up.

The paper is organized as follows. In section \ref{sec:2} we discuss
perturbative aspects of the Polyakov loop and the use of dimensional
reduction to attempt the calculation beyond NLO. In
section \ref{sec:3} we show that the presence of condensates introduce
a power-like pattern in the logarithm of the Polyakov loop expectation
value. In section \ref{sec:4} we analyze the lattice data and show
that they are fairly well described as a composition of perturbative
plus condensate contributions. Finally, in section \ref{sec:5} we
summarize our conclusions.

\section{The perturbative Polyakov loop}
\label{sec:2}

\subsection{Perturbative results}

The (expectation value of the) Polyakov loop is defined as
\begin{equation}
L(T)  = \left\langle \frac{1}{N_c}\tr\, {\bf P} \left( 
e^{i g\int_0^{1/T} d x_0 A_0 (\bfx , x_0) }\right) \right\rangle
\label{eq:1}
\end{equation}
where $\langle~\rangle$ denotes vacuum expectation value, $\tr$ is the
(fundamental) color trace, and ${\bf P} $ denotes path ordering. $A_0$
is the gluon field in the (Euclidean) time direction, $A_0 = \sum T_a
A_{0,a}$, $T_a$ being the Hermitian generators of the SU$(N_c)$ Lie
algebra in the fundamental representation, with the standard
normalization $\tr(T_aT_b)=\delta_{ab}/2$.

As a composite  operator the Polyakov loop is subject to
renormalization. The multiplicative renormalizability of the Polyakov
loop was established in
Refs. \cite{Polyakov:1980ca,Arefeva:1980zd,Dotsenko:1979wb,Gervais:1979fv}
in the context of perturbation theory. Gava and Jengo
\cite{Gava:1981qd} addressed the perturbative computation of $L(T)$ in
pure gluodynamics. The calculation was carried out to NLO, which
corresponds to ${\cal O}(g^4)$, using dimensional regularization and
in the Landau gauge. The result is of course gauge
invariant. Explicitly,
\begin{equation}
L(T)= 1+\frac{1}{16 \pi}\frac{N_c^2-1}{N_c}g^2\frac{m_D}{T} +
\frac{N_c^2-1}{32\pi^2}g^4\left(\log\frac{m_D}{2T}+\frac{3}{4}\right)
+{\cal O}(g^5) \,.
\label{eq:2}
\end{equation}
Here $m_D$ is the Debye mass, which controls the screening of
chromoelectric modes in the plasma. To one loop
\cite{Nadkarni:1983kb}
\begin{equation}
m_D=gT(N_c/3+N_f/6)^{1/2}\,,
\end{equation}
$N_c$ being the number of colors and $N_f$ the number of flavors, to
account for dynamical quarks. The coupling constant $g$ runs with the
temperature following the standard renormalization group analysis and
one expects (\ref{eq:2}) to hold for high enough temperature.
Remarkably, $L(T)$ turns out to be larger than unity implying that the
renormalized Polyakov loop is not a unimodular matrix. Note that $m_D$
contains a $g$ and so the first non trivial contribution to $L$ is
${\cal O}(g^3)$, due to the infrared structure of the theory, rather
than the naively expected ${\cal O}(g^2)$. Note also that the
perturbative result (as well as $m_D$) has a well defined large $N_c$
limit, with 't Hooft prescription of keeping $g^2N_c$ fixed.

\subsection{Dimensional reduction}

The result just quoted is rather old yet no higher order computations
are presently available. Most efforts in perturbative high temperature
QCD have been addressed to obtain the pressure and only recently
such computations have been taken to their highest possible
perturbative order \cite{Kajantie:2003ax}, using dimensional reduction
ideas
\cite{Ginsparg:1980ef,Appelquist:1981vg,Nadkarni:1983kb,Braaten:1996jr,Shaposhnikov:1996th}.
In order to subsequently include possible contributions from
condensates, we will presently reproduce the lowest order perturbative
result for $L(T)$ using the dimensional reduction approach. In
addition this will allow us to discuss properties of higher order
perturbative contributions to $L(T)$.

The starting point is the Euclidean QCD action
($D_\mu=\partial_\mu-ig_0A_\mu$, $F_{\mu\nu}=ig_0^{-1}[D_\mu,D_\nu]$,
$N_f$ massless fermions)
\begin{equation}
{\cal L}_{\text{QCD}}=\frac{1}{2}\tr(F_{\mu\nu}^2)+\bar\psi\thru{D}\psi
+{\cal L}_{\text{gf+gh+ct}} \,,
\end{equation}
where ${\cal L}_{\text{gf+gh+ct}}$ accounts for gauge fixing and ghost
terms as well as the counterterms for renormalization. Next, one
proceeds to integrate out the fermionic modes and all non stationary
gluon modes, which become very heavy at high temperature. This results
in an effective theory for the remaining stationary (time-independent)
gluon modes $A_\mu(\bfx)$, described by a three dimensional action
$\int d^3x\, {\cal L}_3(\bfx)$. To one loop and in the Landau gauge
one obtains
\cite{Huang:1994cu,Chapman:1994vk,Shaposhnikov:1996th,Megias:2003ui}
\begin{eqnarray}
T{\cal L}_3(\bfx) &=&
m_D^2\tr(A_0^2)
+\frac{g^4(\mu)}{4\pi^2 }(\tr(A_0^2))^2
+\frac{g^4(\mu)}{12\pi^2 }(N_c-N_f)\tr(A_0^4)
\nonumber\\
&& +\frac{g^2(\mu)}{g_E^2(T)}\tr([D_i,A_0]^2)
+\frac{g^2(\mu)}{g_M^2(T)}\frac{1}{2}\tr(F_{ij}^2)
+T\delta {\cal L}_3
\label{eq:4}
\end{eqnarray}
where $g(\mu)$ is the running coupling constant in the
$\overline{\text{MS}}$ scheme (to be used in the Debye mass and in the
Polyakov loop formula too)
\begin{eqnarray}
\frac{1}{ g^2(\mu)} &=& 
2\beta_0\log(\mu/\Lambda_{\overline{\text{MS}}}) \,,\qquad
%\nonumber \\
\beta_0 = (11N_c/3-2N_f/3)/(4\pi)^2
\label{eq:6}
\end{eqnarray}
and
\begin{eqnarray}
\frac{1}{ g_E^2(T)} &=& \frac{1}{ g^2(\mu)}
-2\beta_0(\log(\mu/4\pi T)+\gamma_E)
+\frac{1}{3(4\pi)^2}\left(N_c+8N_f\left(\log 2-1/4\right)\right) \,,
\nonumber \\
\frac{1}{g_M^2(T)} &=& \frac{1}{g^2(\mu)}
-2\beta_0(\log(\mu/4\pi T)+\gamma_E) 
+\frac{1}{3(4\pi)^2}\left(-N_c+8N_f\log 2\right) \,.
\end{eqnarray}
The remainder $\delta {\cal L}_3$ contains operators of mass dimension
6 and higher. In addition there are higher loop terms and constant
(field independent) terms which would be relevant for the pressure.
(Note that $g_E$ and $g_M$ are not to be confused with the coupling
constants under the same name appearing, e.g., in
\cite{Kajantie:2002wa}.)

At lowest order we will only need the mass term and the kinetic energy
term of the chromoelectric field (first and fourth terms respectively
in Eq. (\ref{eq:4})). It will be convenient to work with a rescaled
$A_0$ field equal to $g(\mu)/g_E(T)$ times the $\overline{\text{MS}}$
$A_0$ field. To all effects, including the Debye mass and the Polyakov
loop formula which depends on the product $gA_0$, this is equivalent
to using the new $A_0$ field together with $g_E(T)$ as coupling
constant. The latter will be denoted $g(T)$ or just $g$  from now on,
\begin{eqnarray}
{\cal L}_3(\bfx) &=& \frac{m_D^2}{T}\tr(A_0^2)
+\frac{1}{T}\tr([D_i,A_0]^2) +\cdots\,,
\label{eq:8}
 \\
\frac{1}{ g^2(T)} &=& 
2\beta_0\log(T/\Lambda_E) \,,
\nonumber
\end{eqnarray}
with
\begin{equation}
\Lambda_E=
\frac{\Lambda_{\overline{\text{MS}}}}{4\pi}
\exp\left(\gamma_E-\frac{N_c+8N_f(\log 2-1/4)}{22N_c-4N_f}\right) \,.
\label{eq:2.9}
\end{equation}

For computing the QCD pressure one can use any gauge fixing to
integrate the non stationary modes. This is an intermediate step to
carry out the  integration of the remaining
modes. Consequently covariant gauges are often used as they are
computationally simpler. For the Polyakov loop computation the
situation is different; static gauges are preferred to covariant ones
\cite{Nadkarni:1983kb}. A static gauge is one in which $A_0(x)$ is
brought to be time independent by means of a suitable gauge
transformation. In such a gauge Eq. (\ref{eq:1}) becomes
\begin{equation}
L = \frac{1}{N_c}\left\langle \tr\, e^{ig A_0(\bfx)/T } \right\rangle
\,.
\label{eq:10}
\end{equation}
i.e., $L$ depends only on the stationary mode of $A_0$ and so no
information is lost on the Polyakov loop operator if the non
stationary modes are integrated out. Unfortunately, the necessary
perturbative computations of e.g. ${\cal L}_3(\bfx)$, are only
available for covariant gauges. Only in a static gauge the stationary
mode $A_0(\bfx)$ coincides with the logarithm of the Polyakov loop
operator. Therefore, in a covariant gauge the effective action of the
stationary mode is insufficient to recover Polyakov loop expectation
values\footnote{Using the stationary mode in Eq.~(\ref{eq:10}) amounts
to removing the path ordering operator in the definition of the
Polyakov loop, rendering it a gauge dependent
quantity.}. Nevertheless, as we discuss below, the gauge dependence
only affects beyond NLO and the two coefficients in Eq. (\ref{eq:2})
are reproduced using the formulas in, for instance,
\cite{Kajantie:2002wa,Kajantie:2003ax} and the method explained in the
next subsection.

Doing a series expansion of $L(T)$ in Eq. (\ref{eq:10}) one gets
\begin{equation}
L(T)=
1
-\frac{g^2}{2 T^2}\frac{1}{N_c}\langle\tr(A_0^2)\rangle
+\frac{g^4}{24 T^4}\frac{1}{N_c}\langle\tr(A_0^4)\rangle
+\cdots \,.
\label{eq:11}
\end{equation}
$\tr(A_0)$ vanish identically while, the other terms of odd order are
assumed to vanish due to the QCD conjugation symmetry, $A_\mu(x)\to
-A^T_\mu(x)$. The leading contribution is then attached to
$\langle\tr(A_0^2)\rangle$. This quantity has dimensions of mass
squared and so it would vanish in a perturbative calculation at zero
temperature. At finite temperature instead it should scale as $T^2$
modulo slowly varying radiative corrections. Let
$D_{00}(\bfk)\delta_{ab}$ denote the momentum space propagator for the
canonically normalized fields $T^{-1/2}A_{0,a}(\bfx)$, then
\begin{equation}
\langle A_{0,a}^2\rangle
=(N_c^2-1)T\int\frac{d^3k}{(2\pi)^3}D_{00}(\bfk) \,.
\label{eq:7}
\end{equation}
To lowest order  the three dimensional propagator is
\begin{equation}
D^{\text{Pert}}_{00}(\bfk)=
\frac{1}{\bfk^2+m_D^2} \,,
\end{equation}
where the upperscript $\text{Pert}$ indicates that it is a perturbative
contribution. When this is inserted in (\ref{eq:7}) it yields (we
apply dimensional regularization rules)
\begin{equation}
\langle A_{0,a}^2\rangle^{\text{Pert}} = -(N_c^2-1)\frac{T m_D}{4\pi}
\end{equation}
This result used in Eq. (\ref{eq:11}) (and using
$\tr(A_0^2)=A_{0,a}^2/2$) reproduces the perturbative value of $L(T)$
to ${\cal O}(g^3)$.

\begin{figure}[tbp]
\begin{center}
\epsfig{figure=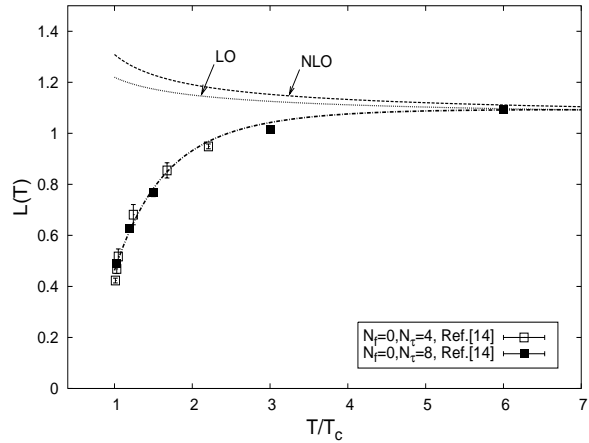,height=6cm,width=8cm}
\end{center}
\caption{The renormalized Polyakov loop versus the temperature, in
 gluodynamics. Lattice data from \cite{Kaczmarek:2002mc}. Perturbative
 LO and NLO results are shown for comparison. The curve follows from a
 fit of the parameter $b$ in Eq.~(\ref{eq:adjust_pert}).}
\label{fig:1}
\end{figure}
In Fig.~\ref{fig:1} we compare the perturbative $L(T)$ in
Eq.~(\ref{eq:2}) with a recent lattice determination of this quantity
in pure gluodynamics and $N_c=3$  \cite{Kaczmarek:2002mc}. As we can
see, in the high temperature region, $T$ about $6\,T_c$, the
$L(T)$-lattice is larger than unity, as predicted by the perturbative
calculation, moreover the numerical value is also consistent with
perturbation theory. The agreement quickly deteriorates as the
critical temperature is approached from above; while the lattice data
moves downwards, to eventually displaying a phase transition, the
perturbative curve increases slightly. As expected, the perturbative
result is slowly varying with temperature, the variation coming from
logarithmic radiative corrections.

\subsection{Higher perturbative orders}

Let us now discuss higher order perturbative contributions to
$L(T)$. The renormalizable pieces of the three dimensional Lagrangian
are of the form
\begin{equation}
{\cal L}_3^{\text{ren}}=
\frac{1}{2}\tr(F_{ij}^2)+\tr([D_i,\A_0]^2)+m^2\tr(\A_0^2)
+\lambda_1(\tr(\A_0^2))^2+\lambda_2\tr(\A_0^4)
\end{equation}
with $\A_0\sim T^{-1/2}A_0$, $m\sim gT$, and
$\lambda_1\sim\lambda_2\sim g^4T$. In addition,
$D_i=\partial_i-ig_3\A_i$ with $\A_i\sim T^{-1/2}A_i$ and $g_3\sim
T^{1/2}g$. For $N_c=2$ or $N_c=3$ the $\lambda_2$ term is redundant
and one can set $\lambda_2=0$. The vacuum energy density of this
theory, $f(g_3,m,\lambda_1)$, has been computed to four loops in
\cite{Kajantie:2003ax}, with $g_3$, $m$ and $\lambda_1$ as independent
parameters. This allows to compute $\langle A_0^2\rangle$ and $\langle
A_0^4\rangle$ by taking derivatives of $f$ with respect to $m^2$ and
$\lambda_1$ respectively, to obtain a perturbative estimate of the
Polyakov loop. The general structure of the vacuum energy density is
as follows \cite{Kajantie:2003ax}
\begin{equation}
f(g_3,m,\lambda_1)=\sum_{\ell\ge 1}
\sum_{k=0}^{\ell-1} f_{\ell k}\,m^{4-\ell} g_3^{2k}\lambda_1^{\ell-k-1}
\label{eq:16}
\end{equation}
where $\ell$ denotes the number of loops and the coefficients $f_{\ell
k}$ depend logarithmically on $m$. Consequently, for the quantities in
the expansion of $L(T)$ one finds
\begin{eqnarray}
\frac{g^2}{T^2}\langle\tr(A_0^2)\rangle
&\sim& 
\frac{g^2}{T}\frac{\partial f(g_3,m,\lambda_1)}{\partial m^2}
\sim
\sum_{\ell\ge 1}\sum_{n=\ell+2}^{3\ell}g^n \,,
\nonumber \\
\frac{g^4}{T^4}\langle\tr(A_0^4)\rangle
&\sim& 
\frac{g^4}{T^2}\frac{\partial f(g_3,m,\lambda_1)}{\partial \lambda_1}
\sim
\sum_{\ell\ge 2}\sum_{n=\ell+4}^{3\ell}g^n  \,.
\end{eqnarray}
As can be seen from these formulas, the first missing contribution to
$L(T)$ would be ${\cal O}(g^7)$ from $\ell=5$ in the
$\langle\tr(A_0^2)\rangle$ term. The lowest contribution from
$\langle\tr(A_0^4)\rangle$ at 5 loops is ${\cal O}(g^9)$ and that from
$\langle\tr(A_0^6)\rangle$, not available from the computation, would
first start at ${\cal O}(g^9)$ at 3 loops. So in principle, one could
extend the perturbative result for $L(T)$ to ${\cal O}(g^6)$.
Unfortunately, the matching relations which connect $m$, $g_3$ and
$\lambda_1$ to the four dimensional QCD parameters are only available in
covariant gauges for which the relation (\ref{eq:10}) does not
apply. In particular, the ratio $g(\mu)/g_E(T)$ used above is gauge
dependent at ${\cal O}(g^2)$ from two loop contributions, this would
introduce a gauge dependence at ${\cal O}(g^5)$ in $L(T)$.

On the other hand, the non renormalizable terms $\delta{\cal L}_3$
ought to be examined as well to determine to which perturbative order
they start contributing to $L(T)$. The leading such terms are
schematically of the type \cite{Chapman:1994vk,Megias:2003ui}
\begin{equation}
\delta{\cal L}_3=
\frac{g^2}{T^2}\tr([D_i,F_{\mu\nu}]^2)
+ \frac{g^3}{T^{3/2}}\tr(F_{\mu\nu}^3)
+ \frac{g^4}{T}\tr(A_0^2 F_{\mu\nu}^2) \,.
\end{equation}
Using the effective relation $D_i\sim gT$, the first term amounts to
an ${\cal O}(g^4)$ correction to the kinetic energy, so it starts
contributing at ${\cal O}(g^7)$ as a correction to the LO. The other
terms are effectively of higher order.

Numerically the terms ${\cal O}(g^5)+{\cal O}(g^6)$ computed with the
available matching relations do not make a substantial contribution as
they are qualitatively and also quantitatively similar to those in
\cite{Gava:1981qd}. Again the radiative nature of these perturbative
terms produces a rather flat logarithmic dependence with the
temperature in sharp contrast with the lattice data at not too high
temperatures. This reinforces the need of non perturbative effects.

\subsection{Gaussian ansatz}

It is noteworthy that the contribution from $\langle A_0^4\rangle$
starts at ${\cal O}(g^6)$, and so to ${\cal O}(g^5)$ $A_0$ obeys a
Gaussian distribution. That is, to this order one can replace
(\ref{eq:10}) with
\begin{equation}
 L  = \exp\left[-\frac{g^2\langle A_{0,a}^2\rangle}{4N_cT^2}
 \right]
\label{eq:18}
\end{equation}
and so
\begin{equation}
\langle A_{0,a}^2\rangle^{\text{Pert}}=
-\frac{N_c^2-1}{4 \pi} m_D T 
-\frac{N_c(N_c^2-1)}{8\pi^2}g^2T^2
\left(\log\frac{m_D}{2T}+\frac{3}{4}\right)+{\cal
O}(g^3) \,.
\label{eq:2b}
\end{equation}
This formula holds also in the unquenched theory, since to this order
$N_f$ only appears through the Debye mass.

The Gaussian ansatz becomes correct ${\cal O}(g^5)$ at high enough
temperature where the theory becomes weakly interacting due to
asymptotic freedom. Also, it becomes exact in the large $N_c$ limit as
higher order connected expectation values are suppressed by powers of
$1/N_c$. Note that $A_{0,a}^2$ scales as $(N_c^2-1)$ and so $L$ has a
well defined limit with the standard prescription of keeping $g^2 N_c$
finite as $N_c\to\infty$. A Gaussian distribution for the Polyakov
loop has been observed in lattice calculations \cite{Engels:1988if}.
The Gaussian ansatz is in fact equivalent to expanding the
exponential, averaging over color degrees of freedom and finally
invoking the vacuum saturation hypothesis ($\langle A_0^{2k} \rangle =
(2 k-1) !!  \langle A_0^{2} \rangle^k$) routinely applied in QCD sum
rules at zero temperature. In this line, the Wilson loop was discussed
in Ref. \cite{Shifman:1980ui} by using the standard dimension 4 gluon
condensate yielding for small contours a term proportional to the area
squared of the contour. The situation has been revisited in
Ref. \cite{Kondo:2002xn} in the context of dimension 2 condensates
yielding an area law for small contours. This agrees with the
observation in Ref. \cite{Chetyrkin:1998yr} that dimension 2
condensates, effectively would-be tachyonic gluon masses, provide the
short range signature of long range confining forces.

\section{Condensate contributions to the Polyakov loop}
\label{sec:3}

As shown in Fig.~\ref{fig:1} the perturbative contributions to the
Polyakov loop expectation value describe only the region of very high
temperature. This situation is reminiscent of what happens for the
heavy quark-antiquark potential in QCD at zero temperature, as a
function of the quark-antiquark separation. There, perturbation theory
describes well the short distance region, where the theory is weakly
interacting and standard one-gluon exchange produces a Coulomb-like
potential. At larger distances confinement sets in and a linear
potential must be added to account for the lattice data
\cite{Bali:2000gf}. As the potential has dimensions of mass, the
Coulomb piece does not need a dimensionful coefficient. This makes it
allowable in perturbation theory, where $\Lambda_{\text{QCD}}$ can
only appear through logarithmic radiative corrections, as in
Eq. (\ref{eq:6}). On the other hand, the linear confining piece of the
potential requires a dimension two coefficient, the string tension,
which in pure gluodynamics should be $\Lambda_{\text{QCD}}^2$ times a
numerical coefficient. At one loop this implies a dependence
$\exp(-1/\beta_0 g^2(\mu))$, the scale $\mu$ being related to the
quark-antiquark separation $r$. While such contributions are perfectly
possible in QCD, they are clearly beyond any finite order in
perturbative QCD and can only be attained through suitable
resummations of the perturbative series (see e.g.
\cite{Brambilla:1999qa,Sumino:2004em}). It is noteworthy that the non
perturbative dependence on $g$ is not completely arbitrary, namely, it
is such that $\Lambda_{\text{QCD}}$ appears raised to positive integer
powers. This finds a natural explanation from the OPE approach, where
the non perturbative contributions are driven by condensates of
concrete local operators. By dimensional counting, the condensate
contributions carry a corresponding negative power momentum
dependence, so they are subdominant at high momentum as compared to
the purely perturbative terms but become more important at lower
momenta, the lower dimensional operators being the dominant ones. In
this line the confining piece of the zero temperature heavy
quark-antiquark potential has been addressed phenomenologically by
considering the contribution to the gluon propagator of a dimension
two condensate, namely, $\langle A_\mu^2\rangle $ in the Landau gauge
\cite{Lavelle:1988eg}. Just by dimensional counting such term produces
a linearly confining term in the potential \cite{Chetyrkin:1998yr}.

In this work we want to investigate the effect of low dimensional
condensates on the Polyakov loop expectation value. The region of high
temperatures is weakly interacting and so ideas inspired on the high
momentum region of the zero temperature theory might be useful here.
As shown above, at high temperatures, the Polyakov loop is closely
related to the expectation value of $A_0^2$ in a static
gauge. Perturbatively, such quantity necessarily scales as $T^2$, but
non perturbatively a further term proportional to
$\Lambda_{\text{QCD}}^2$ is allowed. In order to account for non
perturbative contributions coming from condensates, we will consider
adding to the propagator new phenomenological pieces driven by
positive mass dimension parameters. Specifically, we consider
\begin{equation}
D_{00}(\bfk)=D^{\text{Pert}}_{00}(\bfk)+D^{\text{NonPert}}_{00}(\bfk)
\label{eq:3.1}
\end{equation}
with the non perturbative term
\begin{equation}
D^{\text{NonPert}}_{00}(\bfk)= \frac{m_G^2}{(\bfk^2+m_D^2)^2} \,.
\end{equation}
Such ansatz parallels those made at zero temperature in the presence
of condensates \cite{Lavelle:1988eg,Chetyrkin:1998yr}. This new piece
produces a non perturbative contribution to $\langle A_0^2\rangle$,
namely,
\begin{equation}
\langle A_{0,a}^2\rangle^{\text{NonPert}} = \frac{(N_c^2-1)T m^2_G}{8\pi m_D} \,.
\end{equation}
If we assume that $m_G$ is temperature independent up to radiative
corrections, the condensate will also be temperature independent,
modulo these corrections. Equivalently, in terms of the condensate
\begin{equation}
D^{\text{NonPert}}_{00}(\bfk)= 
\frac{8\pi}{N_c^2-1}\frac{m_D}{T}
\frac{\langle A_{0,a}^2\rangle^{\text{NonPert}}}
{(\bfk^2+m_D^2)^2} \,.
\end{equation}
Note that a positive condensate $\langle A_{0,a}^2
\rangle^{\text{NonPert}}$
indicates a would-be tachyonic gluon mass $-m_G^2$, as in
\cite{Chetyrkin:1998yr}.

Adding the two contributions to $\langle A_{0,a}^2\rangle$ in
Eq.~(\ref{eq:18}), one obtains
\begin{eqnarray} 
-2\log L = 
\frac{ g^2 \langle A_{0,a}^2 \rangle^{\text{Pert}}
 }{ 2N_c  T^2 } 
+
\frac{ g^2\langle A_{0,a}^2 \rangle^{\text{NonPert}}
 }{ 2N_c  T^2 }
 \label{eq:log_L}
\end{eqnarray} 
The fact that, modulo radiative corrections (including running of the
coupling and anomalous dimensions), $\langle A_{0,a}^2
\rangle^{\text{Pert}}$ scales as $T^2$ while $\langle A_{0,a}^2
\rangle^{\text{NonPert}}$ is temperature independent, suggests rewriting
the previous formula as \cite{Megias:2004hj}
\begin{eqnarray}
- 2 \log L = a + b \left(\frac{T_c}T \right)^2  
\label{eq:adjust}
\end{eqnarray}
where the parameters $a$ and $b$ are expected to have only a weak
temperature dependence. As advertised the non perturbative piece
introduces a power-like dependence in the temperature which is not
present in the perturbative calculation.

\section{Comparison with lattice data} 
\label{sec:4}

\subsection{Results in gluodynamics}

A reliable determination of the renormalized Polyakov loop in lattice
gauge theory has been undertaken only recently in
Ref. \cite{Kaczmarek:2002mc}, for pure gluodynamics and $N_c=3$. This
calculation is, of course, fully non perturbative. These authors
compute the finite temperature correlation function of a heavy
quark-antiquark pair for different separations. The two Polyakov loops
are multiplicatively renormalized by extracting the (temperature
dependent but separation independent) quark self energy in such a way
that at short distances the standard zero temperature quark-antiquark
potential is reproduced. At large separations the (squared)
renormalized Polyakov loop is then obtained. That is, if $P_x$ denotes
the renormalized Polyakov loop operator located at $x$,
\begin{equation}
\langle P_x P_y \rangle =
e^{-c(T)}\langle P^{\text{bare}}_x P^{\text{bare}}_y \rangle
=e^{-F_{\bar{q}q}(r,T)/T} 
\underset{{r\to\infty}}{\longrightarrow} L^2(T) \,.
\label{eq:26}
\end{equation}

%%%%%%%%%%%%%%%%%%%%%%%%%%%%%%%%%%%%%%%%%%%%%%%%%%%%%%%%%%
\begin{figure}[tbp]
\begin{center}
\epsfig{figure=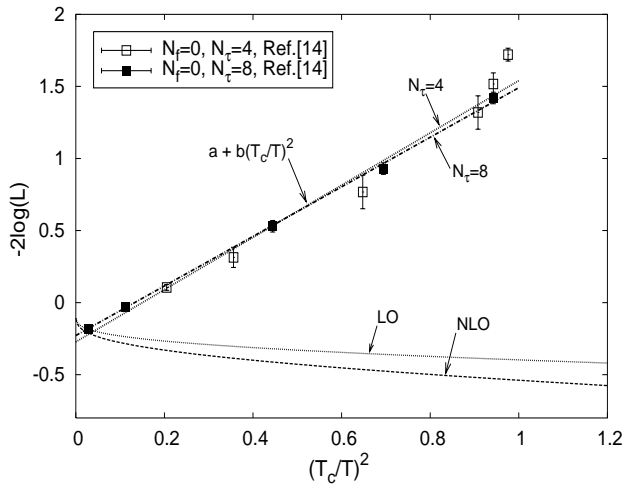,height=6.5cm,width=8.5cm}
\end{center}
\caption{The logarithmic dependence of the renormalized Polyakov loop
 in gluodynamics versus the inverse temperature squared in units of
 the critical temperature. Lattice data from
 \cite{Kaczmarek:2002mc}. The fits use Eq.~(\ref{eq:adjust}) with $a$
 and $b$ adjustable constants and lattice data above $1.03\,T_c$ for
 $N_\tau=4$ and $N_\tau=8$. Purely perturbative LO and NLO results for
 $N_f=0$ are shown for comparison.  }
\label{fig:2}
\end{figure}
%%%%%%%%%%%%%%%%%%%%%%%%%%%%%%%%%%%%%%%%%%%%%%%%%%%%%%%%
Motivated by the pattern in Eq.~(\ref{eq:adjust}), the lattice data
for $-2\log L(T) $ are displayed versus $(T_c/T)^2$ in
Fig.~\ref{fig:2}. As we can see the lattice data follow a
nearly straight line. This pattern is clearly distinguishable from the
much flatter dependence predicted by the perturbative calculation, and
unequivocally shows a temperature power correction characteristic of a
dimension 2 condensate.

Identification of (\ref{eq:adjust}) with the formula
(\ref{eq:log_L}) yields the relations
\begin{eqnarray}
a &=& 
-\frac{1}{8 \pi}\frac{N_c^2-1}{N_c}g^2\frac{m_D}{T} -
\frac{N_c^2-1}{16\pi^2}g^4\left(\log\frac{m_D}{2T}+\frac{3}{4}\right)
+{\cal O}(g^5) 
\,,  \label{eq:apert}\\
g^2 \langle A_{0,a}^2 \rangle^{\text{NonPert}} &=& 2 N_c T_c^2 b\,.
\label{eq:cpert}
\end{eqnarray}
A fit of the lattice data of the form 
\begin{eqnarray}
-2 \log{L} =a^{\text{NLO}}+ b \left(\frac{T_c}{T}\right)^2 
 \label{eq:adjust_pert}
\end{eqnarray}
with the perturbative value of $a$ to NLO and $b$ as a free constant
parameter, yields
\begin{eqnarray}
b = \left\{
\begin{matrix}
      2.20(6)\,, \cr
      2.14(4)\,, 
\end{matrix}
\right.
\qquad
\chi^2/{\rm DOF} = \left\{
\begin{matrix}
      0.75 \,, && N_\tau=4\,,  \cr
      1.43\,, && N_\tau=8\,.  \cr  
\end{matrix} \right.
\label{eq:a_NLO}
\end{eqnarray}
This corresponds to the following value for the condensate
\begin{eqnarray}
g^2 \langle A_{0,a}^2 \rangle^{\text{NonPert}} = \left\{
\begin{matrix}
      (0.98 \pm 0.02 \, {\rm GeV} )^2 \,, &&   N_\tau=4 \,,  \cr
      (0.97 \pm 0.01 \, {\rm GeV} )^2 \,, &&   N_\tau=8 \,.
\end{matrix}       \right.
\label{eq:condfit1}
\end{eqnarray}
In the fit we include lattice data for temperatures $1.03\,T_c$ or
above. We use $T_c/\Lambda_{\overline{\rm
MS}}=1.14(4)$~\cite{Beinlich:1997ia,Bali:2000gf}, and
$T_c=270(2)\;{\rm MeV}$ ~\cite{Beinlich:1997ia}. Throughout this
section we use the running coupling constant obtained from the beta
function to three loops and $\Lambda_E$ in Eq.~(\ref{eq:2.9}) as scale
parameter. Assuming that the difference between the two lattice
results is entirely due to finite cutoff effects, and assuming further
that the corresponding leading effect goes as $1/N_\tau$, we obtain
the estimate $(0.95(4) \, {\rm GeV} )^2$ for $g^2 \langle A_{0,a}^2
\rangle^{\rm NonPert}$ in the continuum limit.

We have also considered a fit of the lattice data with both $a$ and
$b$ treated as free constant parameters. This produces
\begin{eqnarray}
a = \left\{
\begin{matrix}
      -0.27(5) \,, \cr
      -0.23(1) \,,
\end{matrix}
\right.
\qquad
b = \left\{
\begin{matrix}
      1.81(13) \,, \cr
      1.72(5) \,, 
\end{matrix}
\right.
\qquad
\chi^2/{\rm DOF} = \left\{
\begin{matrix}
      1.07\,, && N_\tau=4 \,, \cr
      0.45\,, && N_\tau=8\,.
\end{matrix} \right.
\label{eq:a_cte} 
\end{eqnarray} 
The values of $\chi^2 /{\rm DOF}$ are slightly better than the
NLO prediction of $a$. Obviously the identification of $a$ with the
perturbative result will work better at high temperatures.  Using
Eq.~(\ref{eq:apert}) we obtain for the highest temperature $6\,T_c $
\begin{eqnarray}
a^{\text{NLO}} = -0.22(1)\qquad (T= 6\,T_c) \,,
\end{eqnarray}
in qualitative agreement with the fitted values. Note that for this
temperature the non perturbative power correction does contribute at
the few percent level. For lower temperatures the NLO perturbative
result evolves faster than the fit suggests. At this level of accuracy
one should also take into account logarithmic corrections to the value
of the condensate and eventually some anomalous dimension correction
to the condensate. The present data do not allow a clean extraction of
such fine details. The average value we get for the condensate with
constant $a$ is
\begin{eqnarray}
g^2 \langle A_{0,a}^2 \rangle^{\text{NonPert}} = \left\{
\begin{matrix}
      (0.89 \pm 0.03 \, {\rm GeV} )^2 \,, &&   N_\tau=4 \,,  \cr
      (0.87 \pm 0.02 \, {\rm GeV} )^2 \,, &&   N_\tau=8 \,,
\end{matrix}       \right.
\label{eq:condfit2}
\end{eqnarray}
a little lower than before. The corresponding continuum limit estimate
results in $g^2 \langle A_{0,a}^2 \rangle^{\rm NonPert} = (0.84(6) \, {\rm
GeV} )^2$.

We have attempted to determine the coefficient of a possible $1/T^4$
correction, appending formula~(\ref{eq:adjust}) with a term
$c(T_c/T)^4$. When we fit the lattice data for $N_\tau=8$, this
results in
\begin{equation}
b=2.18(20)\,, \quad c=-0.04\pm 0.24\,,
\end{equation}
with $\chi^2/{\rm DOF}=1.89$, where we have considered the
perturbative value of $a$ to NLO, and
\begin{equation}
a=-0.22(2)\,, \quad b=1.61(24)\,, \quad c = 0.13\pm 0.28\,, 
\end{equation}
with $\chi^2/{\rm DOF}=0.42$, if we treat $a$ as a free constant. The
value of $c$ is compatible with zero in any case, and the errors
overlap with central values for $a$ and $b$ of Eqs.~(\ref{eq:a_NLO})
and~(\ref{eq:a_cte}) respectively. More accurate data are desirable in
order to identify contributions from condensates of dimension 4.

It is noteworthy that a fit to the data completely excludes the
existence of a term of the form $1/T$ in $\log(L(T))$. Such term would
not have a theoretical basis, as no dimension one condensate
exists. However, as noted by the authors of \cite{Kaczmarek:2002mc},
there is a ambiguity in their procedure, which corresponds to adding a
constant to the zero temperature quark-antiquark potential. Such
ambiguity translates into an additive ambiguity in $F_{\bar{q}q}(r,T)$
in Eq.~(\ref{eq:26}), which would give rise a term of the type $1/T$
in $\log(L(T))$. The absence of such term indicates a preference for
the Cornell prescription adopted in \cite{Kaczmarek:2002mc}, namely,
in $V_{\bar{q}q}(r)\sim v_0/r+v_1+v_2 r$ to choose $v_1=0$
\cite{Zantow:2003uh}.

\subsection{Relation with zero temperature condensates}

Although our determination is based on a static gauge, it is tempting
to compare with the zero temperature condensate $g^2 \langle
A_{\mu,a}^2 \rangle$ obtained in the Landau gauge in quenched
QCD. There, one obtains from the gluon propagator $ (2.4 \pm 0.6\,
\text{GeV} )^2 $ \cite{Boucaud:2001st}, from the symmetric three-gluon
vertex $ (3.6 \pm 1.2\, \text{GeV} )^2 $ \cite{Boucaud:2001st}, and
from the tail of the quark propagator $ (2.1 \pm 0.1\, \text{GeV} )^2
$ \cite{RuizArriola:2004en} and $ (3.0-3.4\,\text{GeV} )^2 $
\cite{Boucaud:2005rm}. At zero temperature all Lorentz components are
sampled suggesting a conversion factor of 4 from $g^2 \langle
A_{\mu,a}^2 \rangle$ to $g^2 \langle A_{0,a}^2 \rangle$, but according
to \cite{Lavelle:1988eg}, in the Landau gauge the total condensate
scales as $D-1$, $D$ being the Euclidean space dimension, suggesting
instead a conversion factor of 3. Within the uncertainties of the
lattice data as well as the theoretical ambiguities, the agreement is
remarkable, as the two quenched results refer to different
temperatures and gauges. Finite temperature results for the pressure
in pure gluodynamics~\cite{Kajantie:2000iz,Kajantie:2002pu} yield a
value $(0.93(7)\, {\rm GeV})^2$ for $g^2 \langle A_{0,a}^2
\rangle^{\rm NonPert}$, indicating an overall coherent
picture.\footnote{This value has been obtained from lattice data shown
in figure 2 of Ref.~\cite{Kajantie:2000iz}, and also from figure 1 of
Ref.~\cite{Kajantie:2002pu}, in the temperature region used in our
fits.}

\subsection{Unquenched results}

The renormalized Polyakov loop has also been computed in the
unquenched case, using the technique described above, in
Ref.~\cite{Kaczmarek:2005ui} for two flavor QCD.
%%%%%%%%%%%%%%%%%%%%%%%%%%%%%%%%%%%%%%%%%%%%
\begin{figure}[tbp]
\begin{center}
\epsfig{figure=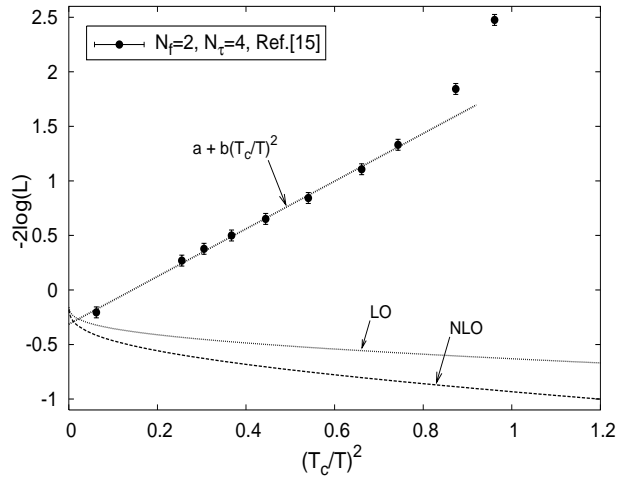,height=6.5cm,width=8.5cm}
\end{center}
\caption{The logarithmic dependence of the renormalized Polyakov loop
 in unquenched QCD with two flavors versus the inverse temperature
 squared in units of the critical temperature. Lattice data from
 \cite{Kaczmarek:2005ui}. The fits use Eq.~(\ref{eq:adjust}) with $a$
 and $b$ adjustable constants and data above $1.15\,T_c$. Purely
 perturbative LO and NLO results for $N_f=2$ are shown for comparison.
 }
\label{fig:3}
\end{figure}
%%%%%%%%%%%%%%%%%%%%%%%%%%%%%%%%%%%%%%%%%%%%
The lattice data are shown in Fig.~\ref{fig:3}, and they corresponds to
$N_\tau=4$. In this case, the data fall onto a straight line for
temperatures $1.15\,T_c$ or above. Closer to the transition temperature
the data start departing from the pattern ~(\ref{eq:adjust}),
indicating the need of a richer description as the
transition is approached from above. A fit to the data above
$1.15\,T_c$ using $a^{\text{NLO}}$ yields
\begin{equation}
b = 2.99(12) \,, 
\quad
 g^2 \langle A_{0,a}^2 \rangle^{\text{NonPert}}= 
(0.86\pm 0.02 \, {\rm GeV} )^2 \,, 
\end{equation}
with $\chi^2/\text{DOF}=1.87$.  We have used
$T_c/\Lambda_{\overline{\rm MS}}=0.77(9)$ with $T_c=202(4)\;{\rm
MeV}$~\cite{Karsch:2000ps} and $\Lambda_{\overline{\rm MS}}=261(31)\;
{\rm MeV}$~\cite{Gockeler:2005rv}. The fit has been done with equal
weight to all data points and the value of $\chi^2$ quoted corresponds
to a representative error $\pm 0.05$ in $2\log L(T)$, which similar to
that for the quenched case.

A fit with $a$ and $b$ as free parameters gives
\begin{eqnarray}
a &=& - 0.31(6)     \, , \quad b = 2.19(13) \,,
\quad
 g^2 \langle A_{0,a}^2 \rangle^{\rm NonPert}= 
(0.73 \pm 0.03 \, {\rm GeV} )^2 \,, 
  \label{eq:adj2} 
\end{eqnarray}
with $\chi^2 /\text{DOF} = 0.25$. As in the quenched case, the value of
$a$ is consistent with the perturbative value at high temperature
\begin{eqnarray}
a^{\text{NLO}} = -0.35(2) \qquad (T= 6\,T_c) \,.
\end{eqnarray}

The lattice data show a departure from the linear pattern for
temperatures closer to the transition than $1.15\,T_c$. Such departure
is not well described by adding new condensates of higher dimension
and we have been unable to extract a condensate of dimension 4 from
the data. We quote here the result of appending a term $c(T_c/T)^4$ in
Eq.~(\ref{eq:adjust_pert}). The fit of the data
above $1.0\,T_c$ gives $b=2.44(21)$ and $c=1.07(19)$ with $\chi^2
/\text{DOF} = 12.8$. The coefficients $b$ and $c$ are highly
correlated.

\subsection{Further quenched lattice data}

\begin{figure}[tbp]
\begin{center}
\epsfig{figure=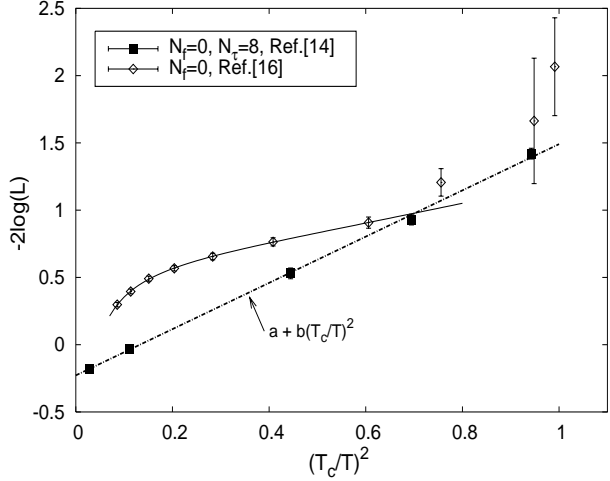,height=6.5cm,width=8.5cm}
\end{center}
\caption{The logarithmic dependence of the renormalized Polyakov loop
 versus the inverse temperature squared in units of the critical
 temperature. Lattice data from
 \cite{Kaczmarek:2002mc,Dumitru:2003hp}. The fits use
 Eq.~(\ref{eq:adjust}) with $a$ and $b$ adjustable constants for the
 first set of data \cite{Kaczmarek:2002mc}, and
 Eq.~(\ref{eq:pis_a_cte}) for the second one \cite{Dumitru:2003hp}. }
\label{fig:4}
\end{figure}

Alternative lattice determinations of the renormalized Polyakov loop
in pure gluodynamics have been addressed more recently in
\cite{Dumitru:2003hp}. These authors follow a different approach as
compared to that in \cite{Kaczmarek:2002mc}. They use single Polyakov
loops which are multiplicatively renormalized by extraction of the
quark selfenergy. The latter is determined by isolating the cutoff
dependent pieces by comparison of different lattice sizes at the same
temperature. Unfortunately the results of both approaches differ
qualitatively, specially for temperatures above $1.3\,T_c$. This is
shown in Fig.~\ref{fig:4} where the two lattice data sets are
compared.

The origin of the discrepancy between the results obtained with the
two approaches is presently not clear, although lattice artifacts, in
particular finite lattice spacing effects, are not completely excluded
in \cite{Dumitru:2003hp} as a possible explanation. (Of course, there
is also the possibility that after closer scrutiny the two definitions
used by the two groups correspond really to different renormalized
operators.)

In our view the results in \cite{Kaczmarek:2002mc} would be the more
reliable ones because the method used is technically simpler and
amenable to tests. Indeed, the authors are able to verify that for
small separations of the two Polyakov loops the standard zero
temperature potential is very accurately reproduced as a function of
$r$ for all temperatures. This is achieved after a single (temperature
dependent) global shift is made, to remove the quark selfenergies;
this is the quantity $c(T)$ in Eq.~(\ref{eq:26}). The contact between
the zero and finite temperature potentials is complete for all
separations between zero and a $T$ dependent radius $r(T)$ related to
the Debye mass, thereby allowing a quite precise determination of the
counterterm $c(T)$ for each temperature. In addition, as noted above,
the calculations are carried out for two different lattice sizes,
$N_\tau=4$ and $N_\tau=8$ (and also $N_\tau=16$ in \cite{Zantow:2003uh}), and
the results for the renormalized Polyakov show very small cutoff
dependence, implying that the continuum limit has been reached. 

The method in \cite{Dumitru:2003hp} is technically more difficult to
implement (quoting the authors, ``In practice, our method is not quite
so trivial'') since it requires comparing different lattice sizes at
the same physical temperature. Also the subtraction of counterterms is
more involved, since, using perturbation theory as guidance, the
analogous of $c(T)$ is expressed as power series of $T$ with
coefficients to be fitted to the bare Polyakov loop data. On the other
hand, from the point of view of the model proposed in the present
work, we expect non perturbative corrections to be negligible at the
highest temperatures considered in the two lattice calculations and
only the data in \cite{Kaczmarek:2002mc} seem to be consistent with
perturbation theory \cite{Gava:1981qd} at those temperatures.

The method in \cite{Dumitru:2003hp} renormalizes the logarithm of the
bare Polyakov loop by using the scheme
\begin{equation}
-\log L^{\text{bare}}(T)=
f^{\text{div}}N_\tau+f^{\text{ren}}+f^{\text{lat}}N_\tau^{-1}
\end{equation}
where $N_\tau$ is the lattice temporal size, and so
$N_\tau=\Lambda/T$, $\Lambda$ being the inverse lattice spacing,
i.e. the lattice cutoff. As said, the data in \cite{Dumitru:2003hp}
deviate from those in \cite{Kaczmarek:2002mc}, and in particular, do
not follow the pattern (\ref{eq:adjust}) for $\log(L)$. Let us make a
speculation assuming that either the removal of the cutoff dependent
pieces has not been complete or that after removal of the those
pieces, finite renormalization terms of the same type as the
subtracted ones remain in the renormalized data of
\cite{Dumitru:2003hp}.\footnote{Of course, one could also ask whether
the result in \cite{Kaczmarek:2002mc} are not contaminated by finite
cutoff effects too, and in particular, whether the linear pattern
displayed in Fig.~\ref{fig:4} is not just the consequence of a huge
cutoff effect of the type $\Lambda^2/T^2$ instead of
$\Lambda_{\text{QCD}}^2/T^2$ as proposed in this work. This
is unlikely, first because the values of the cutoff $\Lambda$ in
\cite{Kaczmarek:2002mc} are much larger than $\Lambda_{\text{QCD}}$
and second, because the renormalized results are consistent for
different lattice sizes, $N_\tau=4$ and $N_\tau=8$.} Specifically, let
us assume that the data follow the pattern
\begin{equation}
- 2 \log L = a^{\text{NLO}} + b \left(\frac{T_c}{T} \right)^2  +
\delta a_{-1}\frac{T_c}{ T}+ \delta a + \delta a_1\frac{T}{T_c} \,.
\label{eq:pis_a_NLO}
\end{equation}
Actually, we find that the data above $1.3\,T_c$ can fairly well be
accounted for by using this pattern. This is shown in
Fig.~\ref{fig:4}.  
%%%%%%%%%%%%%%%%%%%%%%%%%%%%%%%%%%%%%%%%%%%
%%%%%%%%%%%%%%%%%%%%%%%%%%%%%%%%%%%%%%%%%%%%%%%%
Remarkably, the central value of the slope $b$
turns out to be close to that found previously with the other set of
data. However, the best fit has large error bars due to the abundance
of parameters available.
\begin{eqnarray}
\delta a &=& 1.8 \pm 1.8 \,, \qquad\qquad
       b = 1.4 \pm 2.6 \,, \nonumber \\
\delta a_{-1} &=& -1.0 \pm 3.8\,, \qquad\quad
\delta a_1 = -0.29 \pm 0.26 \,,
\end{eqnarray}
with $\chi^2/{\rm DOF}=0.0349$.

Similar remarks apply to the fit
\begin{equation}
- 2 \log L = a + b \left(\frac{T_c}{T} \right)^2  +
\delta a_{-1}\frac{T_c}{ T}+ \delta a + \delta a_1\frac{T}{T_c} \,,
\label{eq:pis_a_cte}
\end{equation}
although in this case $a$ and $\delta a$ cannot be determined
independently. This gives
\begin{eqnarray}
a+\delta a &=& 1.6 \pm 1.8 \,, \qquad\qquad
       b = 1.3 \pm 2.6 \,, \nonumber \\
\delta a_{-1} &=& -1.4 \pm 3.8\,, \qquad\quad
\delta a_1 = -0.28 \pm 0.26  \,,
\end{eqnarray}
with $\chi^2/{\rm DOF}=0.0350$.

We find encouraging that the value of the condensate approximately
agrees using the two different lattice data sets. Nevertheless, this
speculation is not completely conclusive and an agreement between the
results of both lattice groups would be needed before further
consequences could be extracted.

\section{Conclusions} 
\label{sec:5}

There are two main results of our study. First, when suitably
analyzed, the lattice data of the renormalized Polyakov loop above the
deconfinement phase transition show unequivocally the existence of a
non perturbative dimension 2 condensate. Such contributions have not
been considered before but they are in fact dominant and allow to
describe the data in \cite{Kaczmarek:2002mc} down to temperatures {\em
as close to the transition as} $1.03\,T_c$ for pure gluodynamics and
$1.15\,T_c$ for two flavors. Furthermore, the numerical value obtained
from the Polyakov loop is quite consistent with the value of $g^2
\langle A_{0,a}^2 \rangle^{\rm NonPert}$ extracted from the pressure in
gluodynamics.

We have suggested identifying this condensate with the BRST invariant
dimension 2 gluon condensate. Our second finding is that, for pure
gluodynamics, the numerical value of the condensate $\langle
A_{0,a}^2\rangle^{\text{NonPert}}$, defined in a static gauge and extracted
from Polyakov loop data above the deconfinement transition, is
remarkably close to the naive estimate $\langle A_{\mu,a}^2\rangle/4$,
measured at zero temperature and in the Landau gauge. These results
pose the theoretical challenge of establishing the connection outlined
in this paper on a firmer ground. In this light the analogy between
the zero temperature potential and the Polyakov loop noted in the
introduction has been pushed forward in \cite{Megias:2005pe} by
showing that the model in Eq.~(\ref{eq:3.1}) predicts a relation
between the string tension and the slope of the Polyakov loop that is
empirically satisfied.

The simple shape $L^2(T)=e^{-a-b(T_c/T)^2}$ yields $L\to 0$ as $T\to
0$, but does not describe the deconfinement phase transition. The
closest analogy to such transition would be near the inflexion point
of $L^2(T)$, which takes place at a temperature $T_i=(2b/3)^{1/2}T_c$.
This $T_i$ would agree with $T_c$ for a universal geometrical value
$b=3/2$, which not far from the values obtained in this work from
quenched QCD lattice data. Nevertheless, this approximate coincidence
can only be taken as an estimate since the concrete value of the
inflexion point depends on whether $L^2(T)$ or $L(T)$ is used, for
instance. It is noteworthy that the same shape can also be obtained
within the instanton approach at finite temperature %\cite{Megias:2006prep},
along the lines of \cite{Gava:1981dq}. The
relation between instantons and dimension 2 gluon condensates at zero
temperature was suggested in \cite{Hutter:1995sc} and fruitfully
exploited in recent lattice simulations to extract, via cooling
techniques, the infrared behavior of the running coupling constant
\cite{Boucaud:2003xi}. In this regard, it might be rather interesting
to isolate the purely non perturbative instantonic contributions on
the lattice and determine whether, after cooling, the shape
$e^{-(a+b(T_c/T)^2)/2}$ extends also below the phase transition.

\begin{acknowledgments}

We thank K. Kajantie for useful correspondence. This work is supported
in part by funds provided by the Spanish DGI and FEDER funds with
grant no. BFM2002-03218, Junta de Andaluc\'{\i}a grant No. FQM-225 and
EU RTN Contract CT2002-0311 (EURIDICE).
\end{acknowledgments}

%\bibliography{Refs}
%\bibliographystyle{h-physrev3}

\end{document}